\documentstyle[preprint,aps]{revtex}

\newcommand{\be}{\begin{equation}}
\newcommand{\ee}{\end{equation}}

\begin{document}



\title{ Origin of Hall Anomaly in the Mixed State }
\author{ P. Ao }
\address{Department of Theoretical  Physics, Ume\aa{\ }University \\      
         901 87, Ume\aa, Sweden    }

\date{\today}

\maketitle

\abstract{
 There has been a corporative absence of understanding of Hall anomaly data
 in the mixed state in terms of vortex many-body effect and pinning,
 because of the dominant theoretical influence.
 Now D'Anna {\it et al.} [ Phys. Rev. Lett. {\bf 81}, 2530 (1998)
 (cond-mat/9808164)] are brave enough to announce the prominent role
 played by vortex many-body effect and pinning in their 
 interpretation of their own data.
 Here I wish to point out: 
 (1) Indeed the data of  D'Anna {\it et al.} 
     can be explained within an existing
     Hall anomaly theory based on vortex many-body considerations;
 (2) It is not surprising that their data are not consistent with 
    available microscopic Hall anomaly theories, because those theories are 
    mathematically incorrect; and 
 (3) The courage of D'Anna {\it et al}. should be appreciated.  }

\noindent
PACS${\#}$s: 74.25.Fy; 74.60.Ge; 74.72.Bk




The clear correlation between the Hall anomaly and the vortex many-body
effect reported first by D'Anna {\it et al.}\cite{danna}
has not only put all previous independent vortex theories into question,
also required a drastic change in the interpretation of 
the relevant data, where the vortex many-body effect 
has been consistently absent, because of the dominant theoretical
influence.
The purpose of this Comment is to sharpen this situation 
by pointing out:  
(1). Indeed an existing vortex many-body theory for the Hall anomaly 
is consistent with the observation of D'Anna {\it et al.}; and 
(2). Several independent vortex dynamical theories 
are mathematically incorrect. 

Generally, it has been known that independent particle models, 
Drude type models, cannot give 
a proper explanation of the Hall effect in a solid\cite{solid}, 
and that the Hall effect is a result of 
the interplay between the many-body effect 
arising from Fermi statistics and electron-electron interaction
and a background potential, periodic or random.
For the Hall anomaly in the mixed state of a superconductor, 
it is evident that
without fluctuations the vortex interaction dominates:
the formation of the Abrikosov vortex lattice.
One of straightforward ways to account for the collective effects
in the Abrikosov phase is the point-like excitations, such as vortex 
vacancies and interstitials or dislocation pairs, 
though more complicated collective motions may also be relevant. 
In addition to the interaction, the pinning must be strong enough
to prevent the sliding of the whole vortex lattice, 
and makes the defect motions
the dominant contribution to transport properties,
but not too strong to completely destroy the vortex crystalline structure.
This immediately implies that 
at the vortex lattice-melting transition one expects a qualitative change 
of Hall effect, because of the changes in both vortex many-body effect and
pinning.
Furthermore, it obviously implies the pinning plays an important role
in the Hall anomaly phenomena.
Indeed, a theory along this line has been proposed, along 
with several more quantitative prediction which remains to 
be tested by further accurate measurements.\cite{ao}

D'Anna {\it et al.} have made three observations:
 $a)$. The vortex-lattice melting transition affects the Hall behavior;
 $b)$. The Hall conductivity is found to decrease rapidly toward large negative
values below a certain field; and 
 $c)$. Extended, strong pinning defects influence the Hall conductivity in the 
vortex-liquid phase.
Their observation $a)$ is the statement of the vortex many-body effect, and 
$c)$ shows the importance of pinning.
They are evidently consistent with the theory outlined in Ref.\onlinecite{ao}.
The observation $b)$ may be understood as  the nonlinear effects of
generation and motion of vortex lattice defects, such as vacancies, in the 
pinned vortex lattice.
Though a detailed nonlinear theory is not yet available, 
quantitative predictions in the linear regime can be found in 
Ref.\onlinecite{ao}.
An accurate measurement in this regime is desired.  
It should be remarked here that the theory in Ref.\cite{ao} is 
so far the only quantitative Hall anomaly theory based on 
vortex many-body effect.

D'Anna {\it et al.} also noticed their data cannot be explained by 
available independent vortex dynamic theories,
the principle result of their paper.
This is not surprising, because those microscopic theories based on the
relaxation time approximation have been demonstrated 
to be mathematically inconsistent,\cite{ao2}
and the phenomenological argument of using the core electron density
to cancel the Magnus force is not realizable,
because at the phase singular point the associated wavefunction 
amplitude is zero, required by quantum mechanics.\cite{az}
It is unlikely a consistent explanation of D'Anna {\it et al.} data
can be found by an independent vortex dynamics model, {\it i.e.},
a Drude type model for vortices, as suggested in their paper.
After all, we have the successful story of vortex many-body theories in the
explanation of the longitudinal resistivity in the mixed state
up to now,\cite{brandt} 
it is unphysical when coming to the transverse resistivity  
the independent vortex dynamics would dominate.

In conclusion, D'Anna {\it et al.} experiment requires a further development 
of the Hall anomaly theory based on vortex many-body effect and pinning.

This work was support by Swedish NFR.


\begin{thebibliography}{9}
\bibitem{danna}
  G. D'Anna, V. Berseth, L. Forr\'o, A. Erb, and E. Walker,
    Phys. Rev. Lett. {\bf 81}, 2530 (1998).
\bibitem{solid}
  N.W. Ashcroft and N.D. Mermin, {\it Solid State Physics}, pages 15 and 57,
   Saunders College, Philadelphia, 1976.
\bibitem{ao}
    P. Ao, J. Supercond. {\bf 8}, 503 (1995);
         e-print: cond-mat/9702058.
\bibitem{ao2}
   P. Ao,  Phys. Rev. Lett. {\bf 80}, 5025 (1998).
\bibitem{az}
  P. Ao and X.-M. Zhu, 
         Physica {\bf C282-287}, 367 (1997).  
\bibitem{brandt}
  G. Blatter, M.V. Feigel'man, V.B. Geshkenbein, A.I. Larkin, 
  and V.M. Vinokur, Rev. Mod. Phys. {\bf 66}, 1125 (1994);

  E.H. Brandt,  Rep. Prog. Phys. {\bf 58}, 1465 (1995). 
\end{thebibliography}
\end{document}